 \documentclass[final,3p,times]{elsarticle}

\usepackage{amssymb}
\usepackage{lipsum}
\usepackage{amsmath}
\usepackage{amsfonts}

\journal{arXiv}

\begin{document}

\begin{frontmatter}



\title{Radial Stabilization of Magnetic Skyrmions Under Strong External Magnetic Field}


\author[first]{Emir Syahreza Fadhilla}
\author[first]{M Shoufie Ukhtary}
\author[first]{Ardian Nata Atmaja}
\author[second]{Bobby Eka Gunara}
\affiliation[first]{organization={Research Center for Quantum Physics, National Research and Innovation Agency (BRIN)},
            addressline={Kompleks PUSPIPTEK Serpong}, 
            city={Tangerang},
            postcode={15310}, 
            country={Indonesia}}
\affiliation[second]{organization={Theoretical High Energy Physics  Research Division, Institut Teknologi Bandung},
            addressline={Jl. Ganesha 10}, 
            city={Bandung},
            postcode={40132}, 
            country={Indonesia}}

\begin{abstract}
The skyrmion number density, \(q\equiv\vec{n}\cdot\left(\partial_x\vec{n}\times\partial_y\vec{n}\right)/4\pi\), is one of the key quantities that characterizes the topological properties of a magnetic skyrmion. In this work, we propose a model for a two-dimensional magnetic system with Hamiltonian that contains an interaction term proportional to \(q^2\) which preserves inversion symmetry. The proposed \(q^2\) term is also known as the Skyrme term and is a two-dimensional version of the well-known quartic term in models of three-dimensional Hopfions. In contrast with the usual exchange interaction, the \(q^2\) term persists at the strong external magnetic field limit. Using the Landau-Lifshitz-Gilbert equation for micromagnetic calculations, we show that the minimum energy configuration of this model exhibits skyrmion properties. Furthermore, this configuration remains stable under small linear radially symmetric perturbations, and we demonstrate that the total energy of the system is bounded from below, ensuring that it remains above the vacuum energy. This implies a topologically protected configuration. Our model provides a framework for describing skyrmions in materials without broken inversion symmetry, particularly in systems subjected to strong external magnetic fields, where conventional exchange interactions are significantly weaker than the Zeeman effect.
\end{abstract}




\end{frontmatter}




\section{\label{Intro}Introduction}
The magnetic Skyrmion was theoretically proposed as a vortex-like minimum energy configuration of spins in low dimensional systems by some pioneering works \cite{1989JETPBogdanov,BOGDANOV1994255,roessler2006spontaneous,binz2006theory,tewari2006blue}. It is proposed that these structures arise from the competition between the symmetric Heisenberg exchange interactions and the anti-symmetric Dzyaloshinskii-Moriya (DM) interactions, leading to stable, non-collinear spin configurations. The study has garnered significant interest in condensed matter physics due to its exotic properties and potential applications in spintronics \cite{nagaosa2013topological,li2023magnetic}. Furthermore, for a specific two-dimensional space, the Skyrmion structures can be found in various systems, for example in quantum Hall systems \cite{sondhi1993skyrmions}, liquid crystals \cite{Fukuda2011}, acoustics \cite{ge2021observation}, and optical systems \cite{tsesses2018optical,du2019deep,shen2024optical}. 

Recent advancements in theoretical models have provided deeper insights into the formation, dynamics, and interactions of skyrmions. It is known that the Skyrmion's generation mechanism and stability can be improved through the engineering of materials' DM interaction \cite{sharafullin2019dzyaloshinskii,lucassen2020stabilizing}. This can also be used to control Skyrmion's helicity which results in a hybrid magnetic Skyrmion \cite{diaz2016controlling,akhir2024stabilization,liu2024modulation}. These studies show that the DM interaction is central in the dynamics of magnetic Skyrmions, however, the breaking of inversion symmetry in materials is necessary to generate DM interaction \cite{nagaosa2013topological,li2023magnetic}. 

There are also some results which hint that Skyrmion formations are also possible through different types of interactions. One of the most well-known Skyrmionic solutions in models of magnetic systems is the Belavin-Polyakov (BP) Skyrmion which is the exact solution of Hamiltonian that only contains the Heisenberg symmetric exchange interaction term, but this type of Skyrmion is unstable when Zeeman effect is introduced to the system \cite{Polyakov:1975yp,han2017skyrmions}. However, the extended version of this solution is later used as an ansatz to study the Skyrmion profiles of a wider class of Hamiltonian that has interaction terms beyond Heisenberg interaction. These interactions came from various mechanisms, such as anisotropy, external fields, and interaction with lattice structure. It is proposed in \cite{guslienko2015skyrmion} that for materials with weak DM interaction, Skyrmion is stabilized by competition between magnetostatic energy and anisotropic energy. Anisotropy and lattice interaction can also affects the dynamics of Skyrmions by inducing mass that depends on elastic parameters of the material \cite{capic2020skyrmion}. Recent studies also report that the dipolar coupling effect plays a significant role in Skyrmion stabilization in materials with lower DM coupling \cite{utesov2022mean,akhir2024stabilization}. Interestingly, Skyrmion formation is also predicted in systems which preserve inversion symmetry. For example, Skyrmion textures with small integer winding numbers appear as the solution of discrete non-linear Schrodinger equation in two and three-dimensional lattice \cite{kevrekidis2007skyrmion} and in two dimensional honeycomb lattice of interacting fermions \cite{palumbo2015skyrmion}. It is also found that spontaneous breaking of spin rotation in topological spin hall insulators has a vector order parameter with Skyrmion textures \cite{grover2008topological}.

The main quantity which describes the magnetic Skyrmion is the Skyrmion number that is defined by,
\begin{equation}
    Q\equiv\frac{1}{4\pi}\int\vec{n}\cdot\left(\partial_x\vec{n}\times\partial_y\vec{n}\right)d^2x,
\end{equation}
where \(\vec{n}\) is a unit vector corresponding to the orientation of each spin. The Skyrmion number is also known as the topological charge, and its conservation is central to the stability of Skyrmion \cite{nagaosa2013topological,je2020direct}. Such stability is known as topologically supported stability where conservation of topological charge protects the corresponding topological soliton from dispersing into the vacuum \cite{zakharov1986soliton,Manton:2004tk,bravyi2011short}, which ensures robustness against perturbations, making them promising candidates for data storage and logic devices \cite{Kiselev_2011,xu2023reconfigurable}.

The conservation of the Skyrmion number in topologically supported stability implies the existence of a topological current, \(j^\mu\equiv\varepsilon^{\mu\alpha\beta}\vec{n}\cdot\left(\partial_\alpha\vec{n}\times\partial_\beta\vec{n}\right)/8\pi\), \(\mu,\alpha,\beta=t,x,y\), whose temporal component, \(j^t\), corresponds to the Skyrmion number density \(q\equiv\vec{n}\cdot\left(\partial_x\vec{n}\times\partial_y\vec{n}\right)/4\pi\) \cite{Manton:2004tk,han2017skyrmions}. \(\varepsilon^{\mu\alpha\beta}\) is the Levi-Civita tensor in \(1+2\) dimensional spacetime, hence \(\partial_\mu j^\mu=0\) by definition. This implies that \(q\) is indeed a density of conserved quantity with \(j^i=\varepsilon^{ij}\vec{n}\cdot\left(\partial_j\vec{n}\times\partial_t\vec{n}\right)/4\pi\), \(i,j=x,y\), as its current. From this point of view, the role of the Skyrmion number as an invariant of the model becomes more obvious. One of the interesting implications is that the topological current also influences the dynamics of the spins inside the system.

\begin{figure}
    \centering
    \includegraphics[width=0.5\linewidth]{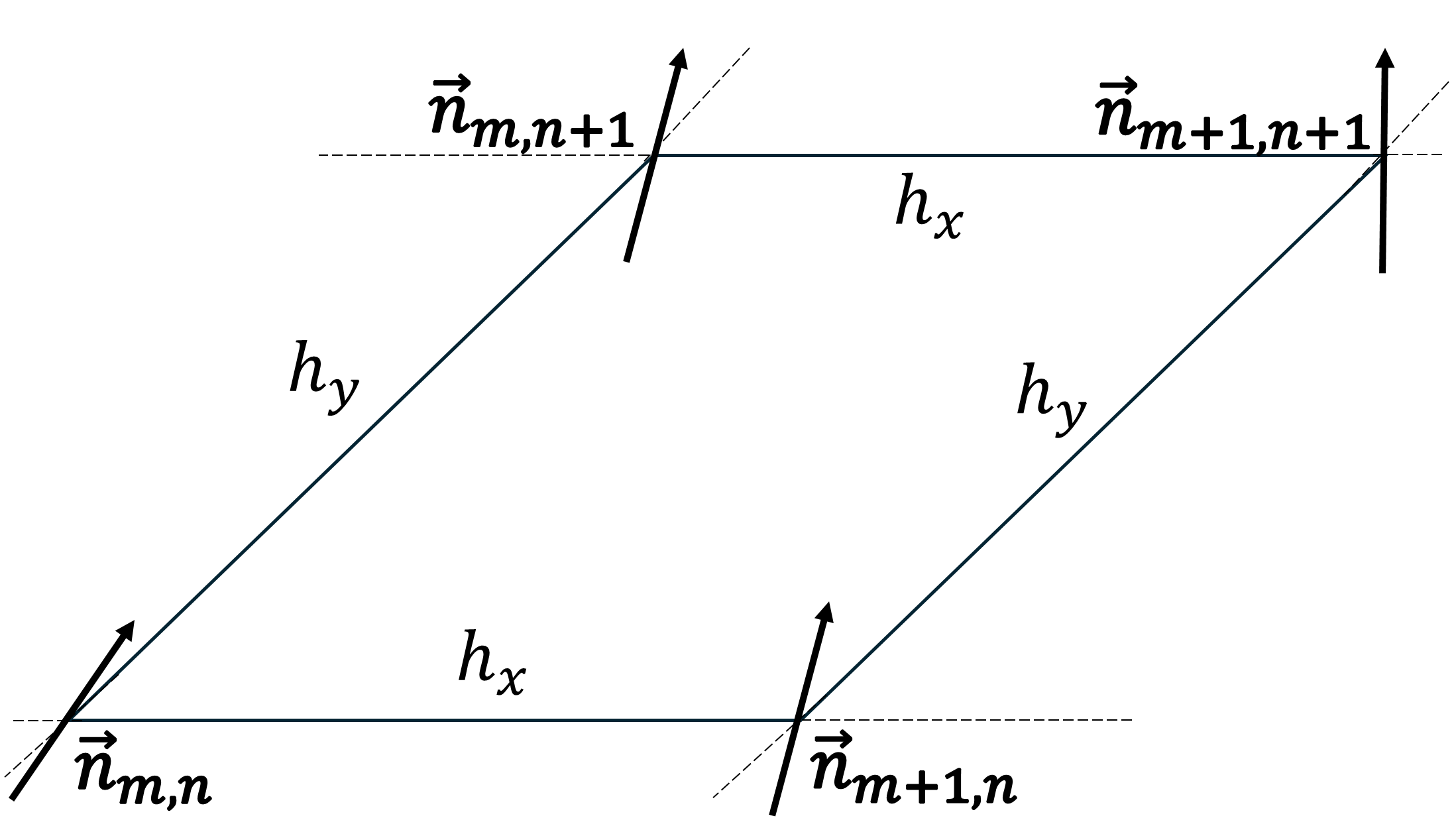}
    \caption{Neighbouring spins in the discrete 2D Skyrme model.}
    \label{fig:array}
\end{figure}
A similar type of influence that depends on \(\vec{n}\cdot\left(\partial_i\vec{n}\times\partial_j\vec{n}\right)\), with  \(i,j=x,y,z\), was first proposed in the three-dimensional classical field theory by Fadeev and Niemi \cite{Faddeev:1976pg} where knot-like solutions are found as the minimal energy configuration of the theory. The knot-like solutions (also known as Hopfions) are predicted to be stable due to their topological property characterized by a topological invariant known as the Hopf number \cite{gladikowski1997static,faddeev1997stable,guslienko2024magnetic}. Micromagnetic simulations show that the formation and collapse of magnetic Hopfions are attributed to the competition between magnetic energies in three-dimensional chiral magnetic systems with certain conditions for the parameters \cite{voinescu2020hopf,rybakov2022magnetic}.  More recent studies further support this by showing that the knot-like structure can be observed in various materials, such as Ir/Co/Pt multilayers \cite{kent2021creation} and FeGe \cite{yu2023realization,zheng2023hopfion}. As such, it is interesting to study whether a quasi-two-dimensional version of \(\vec{n}\cdot\left(\partial_i\vec{n}\times\partial_j\vec{n}\right)\) is also relevant in the study of Skyrmions in two-dimensional systems.

 In the relativistic regime, a topological charge-dependent dynamics has been proposed through the additional term in the field-theoretic Lagrangian that is proportional to \(j^\mu j_\mu\), for example, in high energy physics we have the Bogomol'nyi-Prasad-Sommerfield (BPS) Skyrme model in \(1+3\) dimensional spacetime \cite{Adam:2008uj,Adam:2010fg} that is a good effective model for baryonic particles, and the BPS baby Skyrme model in \(1+2\) spacetime that has been proven to have ferromagnetic properties \cite{Piette:1994mh,Adam:2014xfa}. In the context of relativistic scalar field theory in general \(1+d\) dimensions, the Skyrme field is a map from \(\mathbb{R}^{1,d}\) to \(S^d\) which can be represented by a \(1+d\) dimensional Euclidean vector, \(\phi\in\mathbb{R}^{1+d}\), through embedding with the \(O(d+1)\) sigma model constraint, \(\phi^a\phi^a=1\), \(a=1,\dots,d+1\) \cite{Arthur:1996ia,Fadhilla:2021jiz,atiyah1994introduction}. As such, the special case of the Skyrme model in \(1+2\) dimensional spacetime, with \(\mathbb{R}^2\) spatial submanifold, has a field that is represented by a vector \(\vec{n}\in\mathbb{R}^3\), constrained by \(\vec{n}\cdot\vec{n}=1\). Since the Skyrme field is equivalent to a three-vector, we can see it as any classical vector quantity which preserves its norm throughout the entire space. This property coincides with the assumptions of magnetic models where the elementary spin has a uniform magnitude for every site, hence, the components of the Skyrme field correspond to the orientation of the spin at \((x,y)\). For the static system where \(\vec{n}:\mathbb{R}^{2}\rightarrow S^2\), the correspondence between Skyrmions in field theory and magnetic Skyrmions becomes more straightforward, since both have the same mathematical structure. The main exchange interactions and the Zeeman energy in the Hamiltonian of a magnetic system can be written in terms of \(\vec{n}\)  \cite{1989JETPBogdanov,Barton-Singer:2018dlh}. However, there is a possible term in field theory which has not been widely used in the model of magnetic Skyrmion, namely \(|\varepsilon^{ij}\partial_i\vec{n}\times\partial_j\vec{n}|^2\), that is constructed from Manton's geometrical construction of the Skyrme model \cite{manton1987}. 

According to Manton's work \cite{manton1987,Manton:2024dgh,Manton:2004tk}, the Dirichlet energy of the Skyrme model consists of invariants of strain tensor \(D\equiv \mathcal{J}\mathcal{J}^T\) where \(\mathcal{J}\) is the Jacobi matrix of the harmonic map. The first invariant is \(\text{Tr} [D]\) that is equivalent to the Heisenberg term \(\partial_i\vec
n\cdot
\partial_i\vec{n}\), and the second invariant is \(\det[D]\) that is equivalent to \(|\varepsilon^{ij}\partial_i\vec{n}\times\partial_j\vec{n}|^2\). This second invariant is also proportional to the square of \(q\), due to the fact that \((\vec{n}\cdot(\partial_x\vec{n}\times\partial_y\vec{n}))^2=|\partial_x\vec{n}\times\partial_y\vec{n}|^2\), assuming \(\vec{n}\cdot\vec{n}=1\). As such, we can recast this second invariant as a product of a coupling constant and \(q^2\). This construction of the total energy for the Skyrme model is consistent in arbitrary dimensions \cite{Arthur:1996ia,Brihaye:2017wqa,Fadhilla:2021jiz}. Unlike the usual exchange interactions, which require only two neighbouring spins, the \(q^2\) term requires three spins in a triangular configuration. This makes the term unique to the two-dimensional system, and cannot be found in systems with lower dimensions. The fact that the \(q^2\) can also contribute to the total energy, motivates us to introduce the non-relativistic analogue of \(j^\mu j_\mu\) to the energy of magnetic Skyrmions in condensed matter. At the non-relativistic limit, the temporal scale is much larger than the spatial scale, resulting in \(q=j^0\gg j^i\). Thus, only \(q^2\) is relevant to the energy. We then proceed by discretizing \(q\) on a system with lattice constant \(a\). According to \cite{berg1981definition}, The corresponding discretized version of \(q\) that guarantees integer Skyrmion number is given by 
\begin{equation}\label{DiscreteSkyrmeNumber}
q_{m,n}=\frac{\tan^{-1}\left(\frac{\vec{n}_{m,n}\cdot\left(\vec{n}_{m+1,n}\times\vec{n}_{m,n+1}\right)}{1+\vec{n}_{m,n}\cdot\vec{n}_{m+1,n}+\vec{n}_{m,n}\cdot\vec{n}_{m,n+1}+\vec{n}_{m+1,n}\cdot\vec{n}_{m,n+1}}\right)}{\pi a^2}
\end{equation} 
Here, the site \((m+1,n)\) is located next to \((m,n)\) in the \(\hat{x}\) direction, while the site \((m,n+1)\) is located next to \((m,n)\) in the \(\hat{x}\) direction, as demonstrated in Fig. \ref{fig:array}. We also assume an isotropic lattice with \(h_x=h_y=a\).

\section{Hamiltonian and Minimum Energy Configuration}
\begin{figure}
    \centering
    \includegraphics[width=0.6\linewidth]{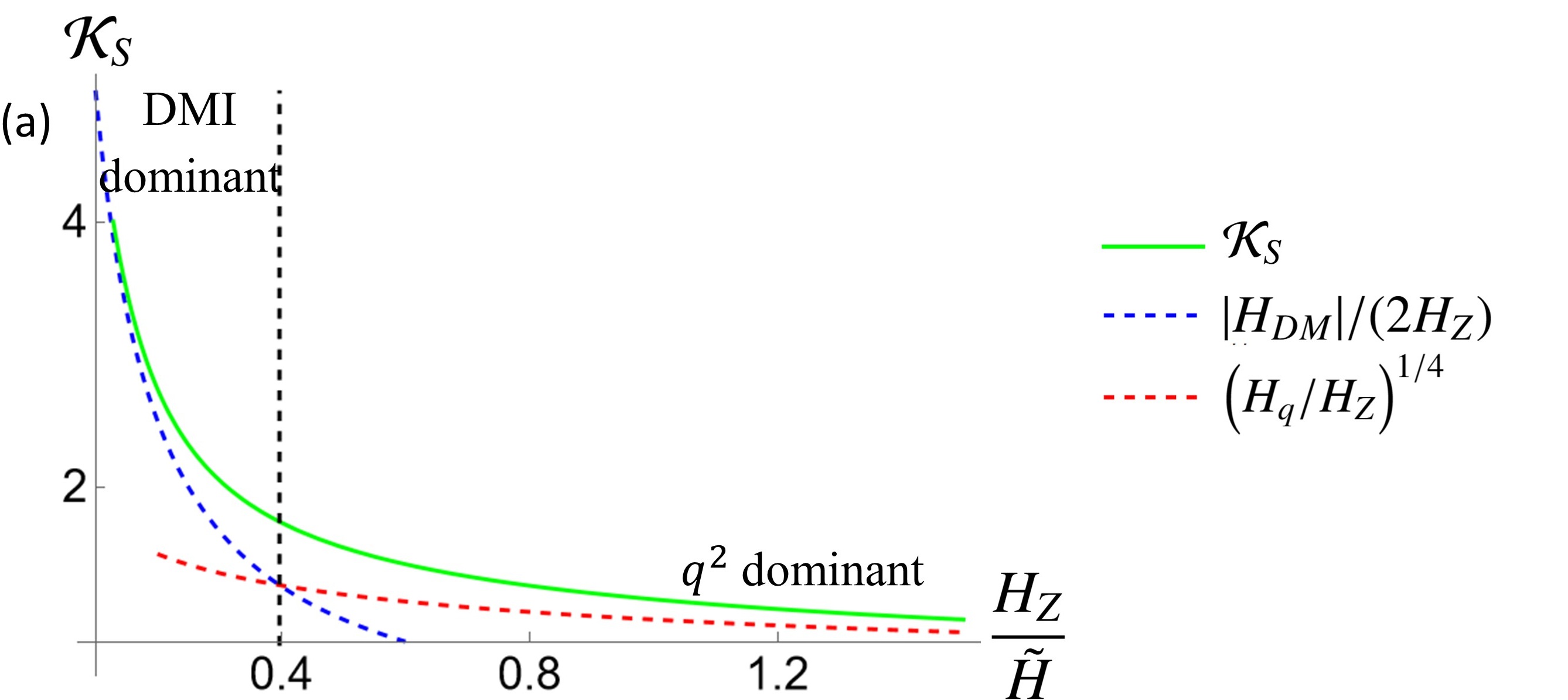}
    \includegraphics[width=0.4\linewidth]{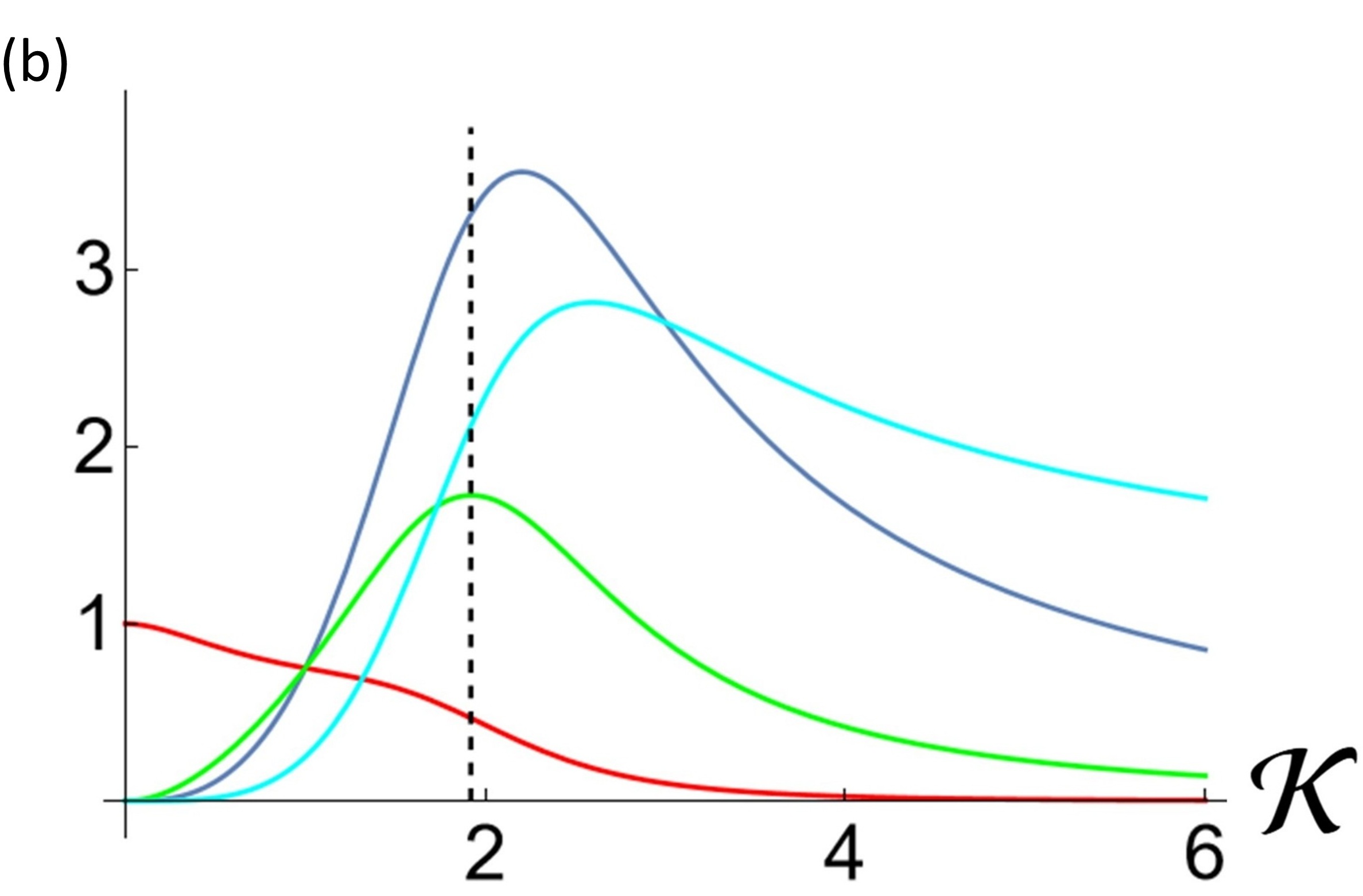}
    \includegraphics[width=0.38\linewidth]{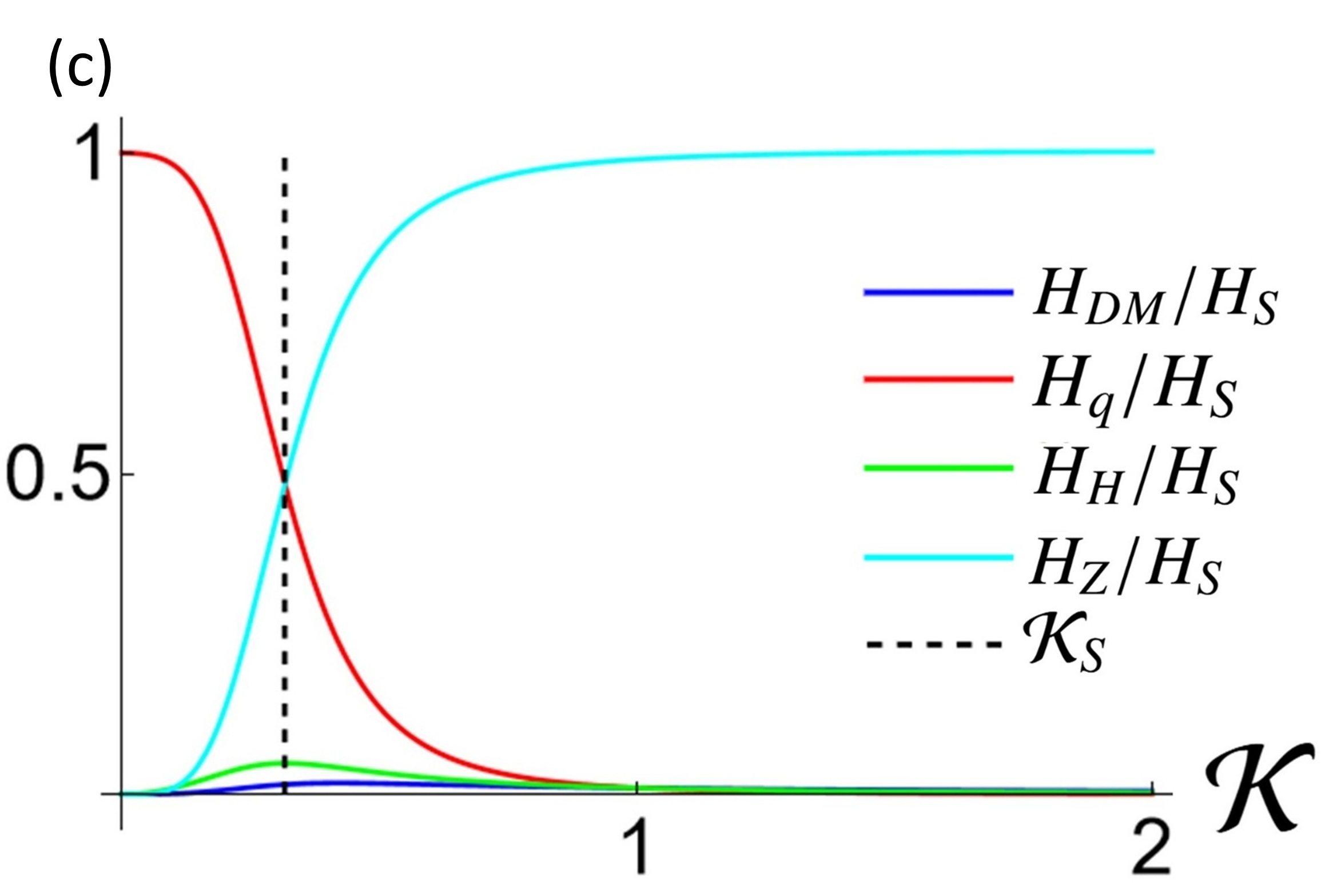}
    \caption{\textbf{(a)} The critical scaling, \(\mathcal{K}_S\) plotted against Zeeman energy, where \(\tilde{H}=H_{DM}=H_q\). We can see that we have two kinds of behaviour, depending on \(H_Z\). \textbf{(b)} The scaled energy ratio, \(|H_i(\mathcal{K})|/H_S(\mathcal{K})\), plotted against the scaling, \(\mathcal{K}\), for \(\tilde{H}/H_Z(1)=3\). \textbf{(c)} \(|H_i(\mathcal{K})|/H_S(\mathcal{K})\), plotted against \(\mathcal{K}\), for \(\tilde{H}/H_Z(1)=0.01\).}
    \label{fig:scalingEnergy}
\end{figure}

The Hamiltonian of the conventional model of magnetic Skyrmions consist of the Zeeman energy \(H_Z\), the Heisenberg symmetric exchange  \(H_H\),  and the DM anti-symmetric exchange \(H_{DM}\).  We can generalize this model by introducing \(H_q\), which is proportional to \(\sum_{m,n}q_{m,n}^2\), such that the total Hamiltonian is \(H_S=H_Z+H_H+H_{DM}+H_q\). All of these terms scale differently under \(a\rightarrow \mathcal{K} a\), \(\mathcal{K}\in\mathbb{R}^+\), and their contribution to the stable size of the Skyrmion depends on this scaling behaviour (See Appendix Section \ref{Discret}). Since \(H_H\) is scale invariant, it does not contribute to the stabilization of Skyrmion's size. Let \(\mathcal{K}_S\) solves \(dH_S(\mathcal{K})/d\mathcal{K}|_{\mathcal{K}_S}=0\). We call \(\mathcal{K}_S\) the critical scaling which represents the size scale of the Skyrmion and we plot \(\mathcal{K}_S\) against variations of \(H_Z(1)/\tilde{H}\) for certain \(\tilde{H}=H_{DM}(1)=H_{q}(1)=H_H(1)\). We can see that we have two regions in Fig. \ref{fig:scalingEnergy} (a), where for small \(H_Z\), \(\mathcal{K}_S\) depends on the competition between \(H_{DM}\) and \(H_Z\), but for large \(H_Z\), \(\mathcal{K}_S\) depends on the competition of \(H_q\) and \(H_Z\) instead. The transition between the two cases happens at approximately \(H_Z\simeq|H_{DM}|^{4/3}/(16H_q)^{1/3}\). We demonstrate this further in Fig. \ref{fig:scalingEnergy} (b) and (c), where we plot the ratio of each rescaled term, \(|H_i(\mathcal{K})|/H_S(\mathcal{K})\), against \(\mathcal{K}\). We demonstrate that for \(\tilde{H}/H_Z(1)=3\) (Fig. \ref{fig:scalingEnergy} (b)) the contribution of \(H_q\) is negligible at \(\mathcal{K}=\mathcal{K}_S\), but for large \(H_Z\) such that \(\tilde{H}/H_Z(1)=0.01\) (Fig. \ref{fig:scalingEnergy} (c)), the only competing terms at \(\mathcal{K}=\mathcal{K}_S\) are \(H_q\) and \(H_Z\) with other terms being negligible. In this limit, the effective magnetic field on \(\vec{n}_{m,n}\) is dominated by the external magnetic field, which implies that other interactions, such as dipole-dipole interaction, are also negligible.

 This work aims to study the role of the \(q^2\) term in the stabilization and to analyze the properties of the resulting skyrmions compared to the conventional models with broken inversion symmetry. In order to do so, we need to isolate the \(q^2\) term such that the stabilization dominantly originates from \(q^2\). According to the scaling behaviour of each term, the critical scale depends only on the competition between \(H_q\) and \(H_Z\) when \(H_Z\) is large, as implied by Fig. \ref{fig:scalingEnergy} (a). Thus, at strong external field limit, we can neglect the other exchange interaction such that the Hamiltonian is effectively given by the sum of \(H_Z\) and \(H_q\), or explicitly,
\begin{eqnarray}\label{Hamiltonian}
    H&=&\kappa_0a^2\left[\sum_{m,n}\left[1-\hat{z}\cdot\vec{n}_{m,n}\right]+32\pi^2\Lambda_2\sum_{m,n}q_{m,n}^2\right].
\end{eqnarray}
\(\kappa_0\) is the density of Zeeman energy from the interaction with a perpendicular magnetic field passing through the plane, and \(\Lambda_2\) is the coupling constant of \(q^2\) term. Such a term is equivalent to the Skyrme term in the classic strong interaction model \cite{Skyrme:1961vq,Skyrme:1962vh} and is a quasi-two-dimensional version of the second term in Fadeev's model of Hopfions. In this limit, the inversion symmetry is recovered, and the remaining possible stabilization of Skyrmion is only through the conservation of Skyrmion number. This setup obeys the Derrick-Hobart theorem (See Appendix Section \ref{Discret} for details).
\begin{figure}
    \centering
    \includegraphics[width=0.35\linewidth]{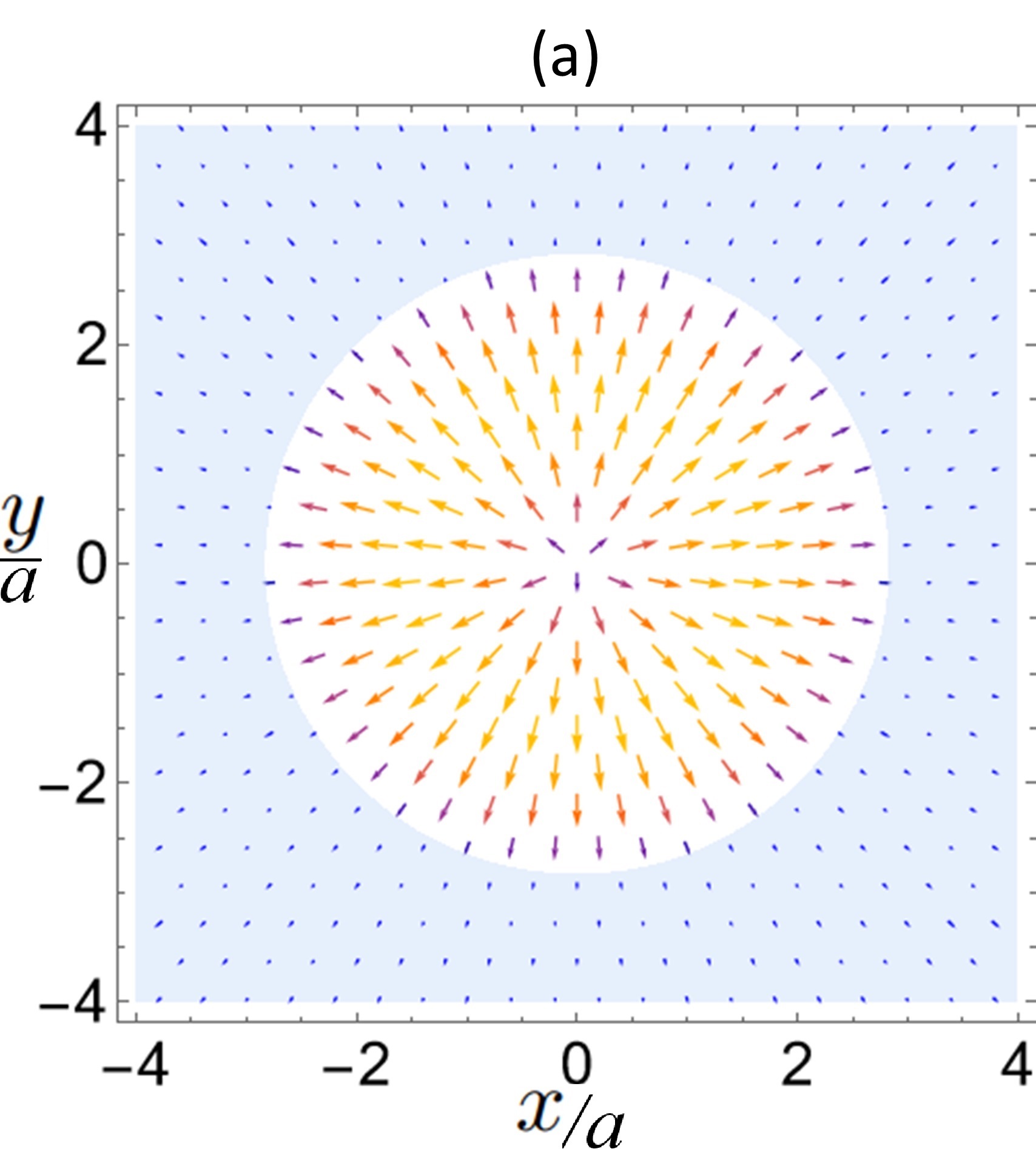}
    \includegraphics[width=0.35\linewidth]{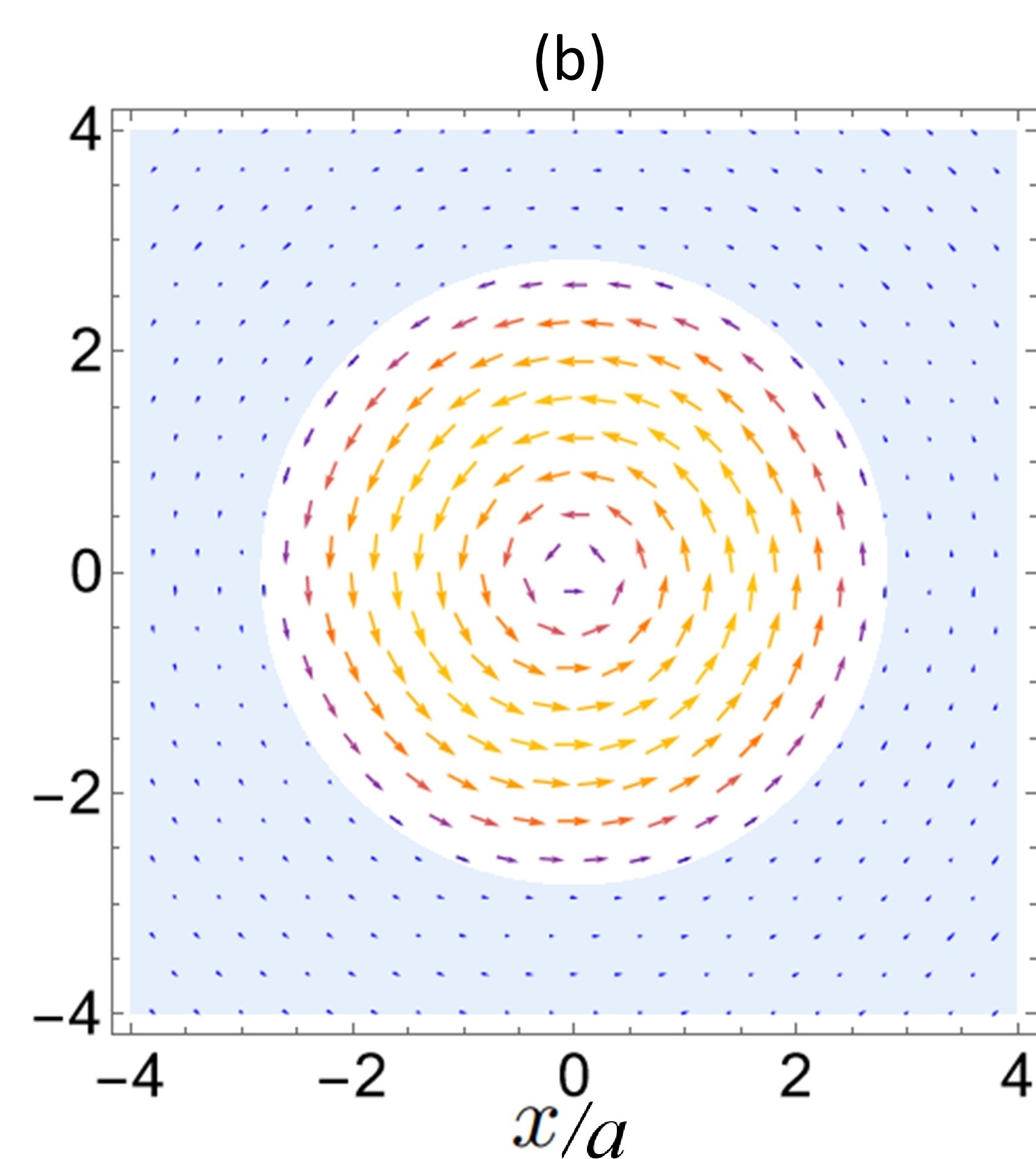}
    \includegraphics[width=0.4\linewidth]{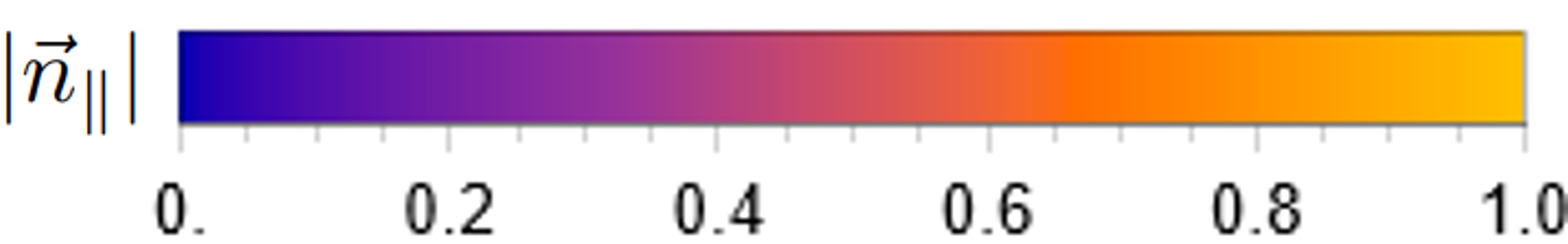}
    \caption{The vector plot of \(\vec{n}_{\parallel}\) on \(z=0\) plane, for Neel-Skyrmion \textbf{(a)} and Bloch-Skyrmion \textbf{(b)}, with \(\Lambda_2/a^4=1\). \(\lambda(r)=0\) in the blue region and \(\lambda(r)\neq0\) in the white region.}
    \label{fig:BPSSkyrmionAbove}
\end{figure}
\subsection{Discrete Limit}
The dynamics of the Skyrmion are assumed to follow the Landau-Lifshitz-Gilbert (LLG) equation that is, for site (\(m,n\)), given by \begin{equation}\label{DiscreteLLG}
    \partial_t\vec{n}_{m,n}=g\left[\vec{n}_{m,n}\times\bar{\delta} H/(a^2\bar{\delta} \vec{n}_{m,n})+\eta \vec{n}_{m,n}\times\partial_t\vec{n}_{m,n}\right].
\end{equation} Here, \(\bar{\delta}\) is the variational operator, \(g>0\) is the gyromagnetic ratio, and \(\eta>0\) is the Gilbert damping constant \cite{han2017skyrmions}. The non-zero Gilbert constant is necessary in the search of the minimum energy configuration through the damping effect. The spin orientation at each sites can be found by solving the Eq. \eqref{DiscreteLLG} up to large \(t\) for
\begin{equation}\label{ansatz}
    \vec{n}_{m,n}=(\sin\theta_{m,n}\cos\varphi_{m,n},\sin\theta_{m,n}\sin\varphi_{m,n},\cos\theta_{m,n}),
\end{equation}
where \((\theta_{m,n},\varphi_{m,n})\equiv(\theta_{m,n}(t),\varphi_{m,n}(t))\) are functions of \(t\) and we approximate \(\theta_{m+1,n+1}\simeq \theta_{m,n}+\Delta^x_{m,n}\theta+\Delta^y_{m,n}\theta\). To simplify the expressions, we define shorthand notations \(\tilde{g}=g/(1-g^2\eta^2)\), \(\tilde{\eta}=g\eta\), and rescale the time \(t\rightarrow t'=\tilde{g} \kappa_0 t\). At large \(t'\), \(\partial_{t'}\vec{n}=0\) and the minimum energy configuration is achieved. At this static steady state, the spin orientations \(\vec{n}_{m,n}\) satisfies
\begin{equation}\label{StaticSolDiscret}
    \vec{n}_{m,n}\times\hat{z}=64\pi^2\Lambda_2q_{m,n}~\vec{n}_{m,n}\times\left(\bar{\delta}q_{m,n}/\bar{\delta}\vec{n}_{m,n}\right)~.
\end{equation}

There are three important regions describing the Skyrmion texture, namely the Skyrmion's center, the Skyrmion's boundary, and the region far away from the center. Let \((m_c,n_c)\) be the center of the Skyrmion. At \(m,n\) far away from \((m_c,n_c)\), the orientation of the spins are almost homogenous, \(\Delta^x_{m,n}\theta=\Delta^y_{m,n}\theta=0\), which leads to \(\vec{n}_{m,n}\times\hat{z}=0\) according to \eqref{StaticSolDiscret}. This leads to \(\vec{n}_{m,n}=\pm\hat{z}\). The \(\vec{n}_{m,n}=\hat{z}\) configuration is more favorable since it minimizes the Zeemann energy, which implies that the spins are parallel to the external magnetic field at regions far from the Skyrmion. On the other hand, the center is characterized by spin that is anti-parallel to the magnetic field, \(\theta_{m_c,n_n}=\pi\). Evaluating \eqref{StaticSolDiscret} at the center gives us \(\Delta^x_{m_c,n_c}\theta=\Delta^y_{m_c,n_c}\theta\) and \(\Delta^x_{m_c,n_c}\varphi=-\Delta^y_{m_c,n_c}\varphi\), which implies that \(\vec{n}\cdot\hat{z}\) is increasing isotropically at the neighbourhood of \((m_c,n_c)\). From these descriptions, \(\vec{n}\cdot\hat{z}\) needs to pass through zero as we trace each site from the center to the far region. This property implies that there exist points where \(\vec{n}\cdot\hat{z}=0\) and the set of these points is the boundary of the Skyrmion. In order to study the profile of the Skyrmion texture in a more detailed manner, we discuss the continuum limit of the model by assuming a large number of spins in the following section. 
\subsection{Continuum Limit}\label{Result} 
In this section, we discuss the properties of a single Skyrmion modelled by the continuum limit of \eqref{Hamiltonian}. The Skyrmion profile we consider here is the simplest single Skyrmion system that assumes axial symmetry and unit Skyrmion number \(Q=1\), whose solutions \eqref{ansatz} satisfy the Belavin-Polyakov ansatz \cite{Polyakov:1975yp},
\begin{equation}\label{BPSol}
    \theta(x,y)=2\tan^{-1}\left(\frac{\lambda}{\sqrt{x^2+y^2}}\right),~~~\varphi(x,y)=\tan^{-1}\left(\frac{y}{x}\right)+\gamma
\end{equation}
Here, \(\gamma\) determines the helicity of the Skyrmion where \(\gamma=0\) and \(\gamma=\pi\) correspond to the Neel Skyrmion and \(\gamma=\pm\pi/2\) correspond to the Bloch Skyrmion. In the standard BP solution, \(\lambda\) is a constant which corresponds to the radius where \(\hat
{z}\) component of \(\vec{n}\) is zero, hence effectively describes the size of the Skyrmion. In this model, we assume that \(\lambda(r)\) varies along the radial coordinate. As such, the profile of the Skyrmion in this model effectively depends only on the function \(\lambda(r)\). The non-constant value of \(\lambda(r)\) implies that \(\lambda\) no longer serves as the size of the Skyrmion, and instead, we define the radial size of the Skyrmion, \(r_s\), as the solution where \(\lambda(r_s)=r_s\) such that \(\hat{z}\cdot\vec{n}|_{r_s}=0\). Substituting Eq. \eqref{BPSol} to \eqref{Hamiltonian}, gives us the continuum limit of the Hamiltonian,
\begin{equation}\label{EnergyBPSModel}
    H\simeq 4\pi\kappa_0\int_0^R\frac{(\lambda/r)^2}{(\lambda/r)^2+1}\left(1+\frac{16\Lambda_2}{r^2\left((\lambda/r)^2+1\right)^3}\left[\frac{d}{dr}\left(\frac{\lambda}{r}\right)\right]^2\right)~r~dr.
\end{equation}
Here, \(R\) is the radius of the magnetic film and the unit for length used in this continuum approximation is the lattice constant, \(a\).

\begin{figure}
    \centering
    \includegraphics[width=0.48\linewidth]{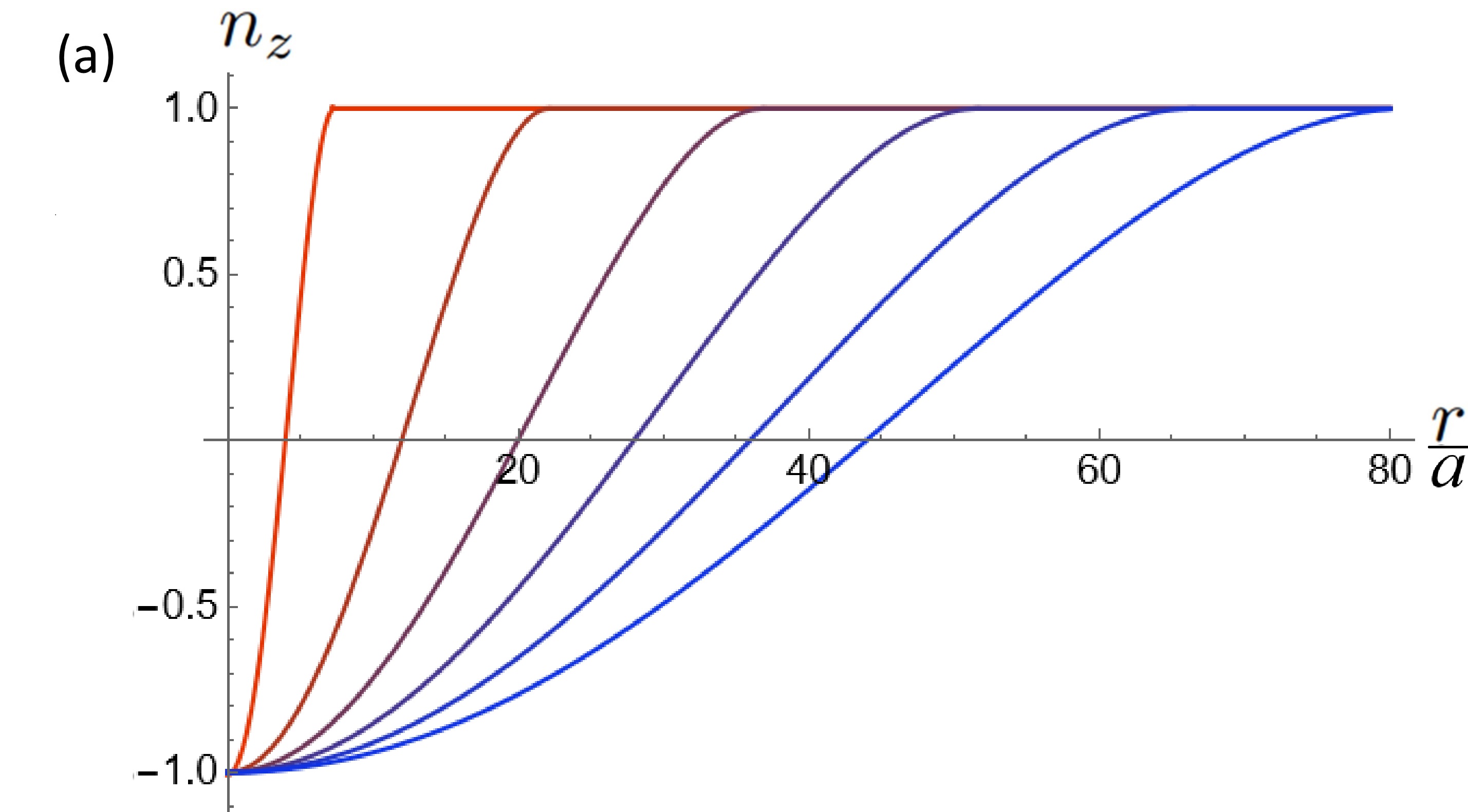}
    \includegraphics[width=0.46\linewidth]{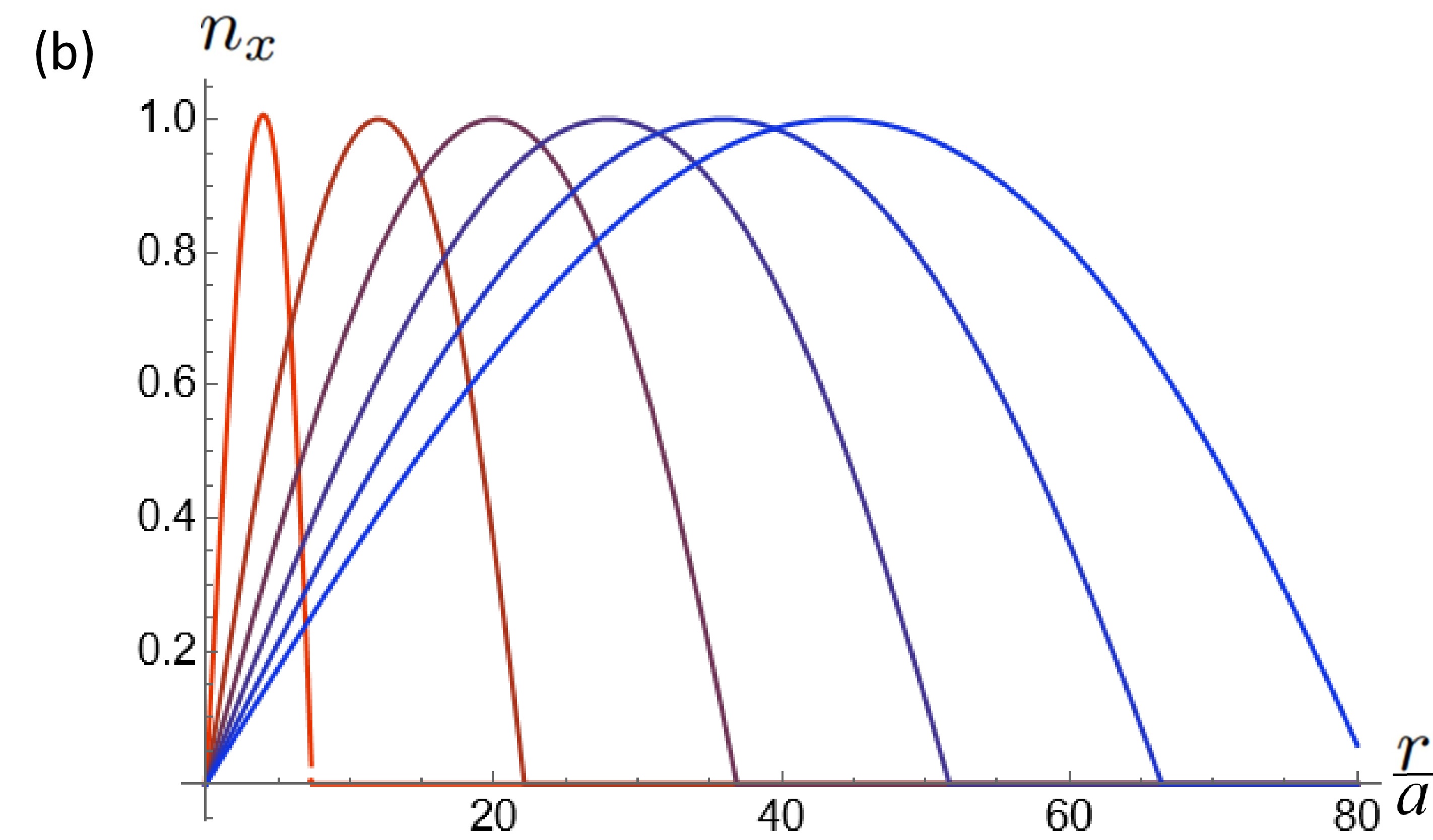}
    \caption{The plot of \(\hat{z}\) \textbf{(a)} and \(\hat{x}\) \textbf{(b)} components of \(\vec{n}\) against \(r\) with variation of \(\Lambda_2^{1/4}/a\) from \(2.6\) (red) to \(28.6\) (blue).}
    \label{fig:nz_LambdaVariation}
\end{figure}

The static solution of the Euler-Lagrange (EL) equations of Eq. \eqref{EnergyBPSModel} is equivalent to the minimum energy configuration which can be found by solving \(\bar{\delta} H/\bar{\delta} \vec{n}=0\). From the EL equation of \(\gamma\), we found that \(\partial_\gamma H=0\), implying that we have degenaracy with respect to the helicity. For the radial profile obeying the EL equation of \(\lambda\), we found that the solution is non-trivial on a finite interval \(r\in[0,2\sqrt2\Lambda_2^{1/4}]\) and, beyond this interval, the solution is trivial, \(\lambda=0\) (See Appendix section B for details)
\begin{equation}\label{ProfileLambda}
    \lambda(r)= \begin{cases} 
      r\frac{8\sqrt{\Lambda_2}-r^2}{\sqrt{64\Lambda_2-\left(8\sqrt{\Lambda_2}-r^2\right)^2}} &~~~,~ 0\leq r\leq 2\sqrt{2}\Lambda^{1/4}_2, \\
      0 &~~~,~ r> 2\sqrt{2}\Lambda^{1/4}_2.
   \end{cases}
\end{equation}
The trivial solution corresponds to the region where the spins are completely parallel to \(\hat{z}\) due to the Zeeman effect. The corresponding vector plot of \(\vec{n}_{\parallel}\), that is \(\vec{n}\) projected onto the \(z=0\) plane is given in Fig. \ref{fig:BPSSkyrmionAbove}, where we can see that both Bloch and Neel-type Skyrmion has the same radial profile. Furthermore, by solving the equation for skyrmion size, \(\lambda(r_s)=r_s\) using the non-trivial solution of \(\lambda(r)\), we find that the size of the skyrmion is proportional to \(\Lambda^{1/4}_2\), namely \(r_s=2\sqrt{2-\sqrt{2}}\Lambda^{1/4}_2\). A typical Skyrmion scale size is on the interval 1 nm - 100 nm \cite{wang2018theory,wu2021size}, which implies that the length-scale of \(\Lambda_2\) should approximately satisfy \(0.18~ \text{nm}^4 \geq \Lambda_2 \geq 65.33 ~\text{nm}^4\) in order to have a realistic size of Skyrmion. As demonstrated in Fig. \ref{fig:nz_LambdaVariation}, larger values of \(\Lambda_2\) results in larger Skyrmion size. Such property is attributed to the fact that the Skyrmion formation in this model comes from the competition between the Zeemann term and the \(q^2\) term in the Hamiltonian, such that if \(\Lambda_2\) is larger then the effect of the external magnetic field becomes weaker and the spins are less likely to become parallel to the \(\hat{z}\) axis. This results in a larger number of spins forming the Skyrmion configuration with \(\hat{z}\cdot \vec{n}<1\). Furthermore, we can see in Fig. \ref{fig:nz_LambdaVariation} that there is a cut-off radius where \(\hat{z}\cdot \vec{n}\) is exactly equal to \(1\) (which also coincides with \(\hat{x}\cdot \vec{n}=0\)). This is the radius where \(\lambda=0\) and all the spins beyond this radius are aligned to the external magnetic field. Such property is, again, contrasting the solution of this model with the usual single Skyrmion where \(\hat{z}\cdot \vec{n}=1\) is achieved at the asymptotic limit \cite{Polyakov:1975yp,1989JETPBogdanov,han2017skyrmions}.
\section{Radial Stability of Skyrmions}\label{Stability}
Substituting the BP ansatz \eqref{BPSol} to the continuous version of the LLG equation \eqref{DiscreteLLG} (with \(\vec{n}_{m,n}\rightarrow\vec{n}(x,y)\)) gives us
\begin{equation}
    \label{LLGEquation}
    \partial_t\vec{n}=\left[1+4\frac{\Lambda_2}{r}\partial_r\left(\frac{1}{r}\cdot\partial_r\left[\hat{z}\cdot\vec{n}\right]\right)\right]\left(\hat{z}\times\vec{n}+\tilde{\eta}\left(I-\vec{n}\otimes\vec{n}\right)\cdot\hat{z}\right),
\end{equation}
where \(I\) is a \(3\times3\) identity matrix and \(\vec{n}\otimes\vec{n}\) is the outer product of \(\vec{n}\) with itself.   

From the static minimum energy configuration found in Section \ref{Result}, we study the stability of this configuration by introducing a perturbation around the static configuration, satisfying \(\lambda(r)\rightarrow\tilde{\lambda}(r)=\lambda(r)+\epsilon\delta\lambda(t,r)\) and \(\lambda\rightarrow\tilde{\gamma}=\gamma+\epsilon\delta\gamma(t)\) with perturbation parameter \(\epsilon\). Since \(\vec{n}\) in Eq. \eqref{BPSol} can be decomposed into \(\vec{n}=R_z(\gamma)\vec{n}_{\gamma=0}\) where \(R_z(\gamma)\) is a rotation matrix around \(\hat{z}\) with rotation angle \(\gamma\), then \(\partial\vec{n}/\partial\gamma=\hat{z}\times\vec{n}\). Thus, the leading order perturbation of \(\vec{n}\) can be expressed as, 
\begin{equation}\label{Tilden}
    \tilde{\vec{n}}=\vec{n}+\epsilon\left(\frac{\partial\vec{n}}{\partial\lambda}\delta\lambda+\hat{z}\times \vec{n}~\delta\gamma\right)\equiv\vec{n}+\epsilon\delta\vec{n},
\end{equation}
We can expand the perturbation of \(\vec{n}\) in terms of perturbation on \(\lambda\) and \(\gamma\) since they are the only quantities defining the profile of the Skyrmion. By substituting the perturbed spin, \(\tilde{\vec{n}}\), from Eq. \eqref{Tilden} to Eq. \eqref{LLGEquation}, and isolating the terms that are linear in \(\epsilon\), we arrive at the first order perturbation equations containing the time-derivatives for each quantity, \(\partial_t\delta\lambda\) and \(\partial_t\delta\gamma\). Following this approach, it is found that the Skyrmion's helicity is neutrally unstable, \begin{equation}
    \partial_t\delta\gamma=0,
\end{equation} which implies that the model \eqref{Hamiltonian} does not have any preference on the Skyrmions' helicity. As such, if we prepare the Skyrmion with helicity \(\gamma_0\) and then perturb the system such that the helicity becomes \(\gamma_0+\delta\gamma\), the system will maintain this perturbed helicity indefinitely. This feature distinguishes our model from the ferromagnetic model with DM term where the helicity can be controlled from the coupling constant of DM interaction \cite{srivastava2018large,akhir2024stabilization}. 

\begin{figure}
    \centering
    \includegraphics[width=0.5\linewidth]{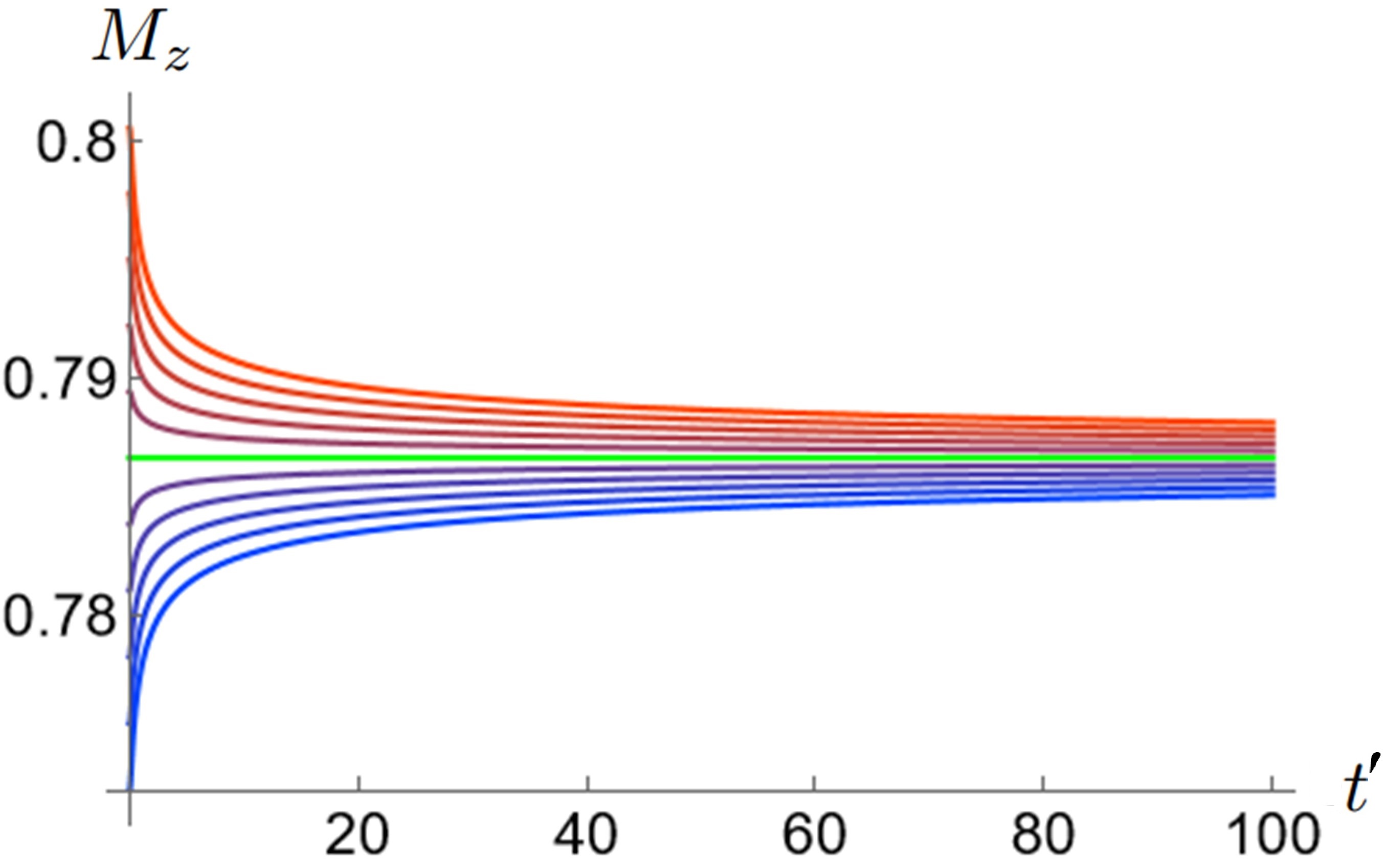}
    \caption{The \(\hat{z}\) component of magnetization, \(\vec{M}\), plotted over time with variation of perturbation magnitude from \(\delta\lambda_0=-0.1\Lambda_2^{1/4}\) (red) up to \(\delta\lambda_0=0.1\Lambda_2^{1/4}\) (blue). The green line shows the magnetization of the unperturbed system.}
    \label{fig:DynamicsMagnetization}
\end{figure}

On the other hand, the dynamics of \(\delta\lambda\) is non-trivial for the case of small perturbation, \(\delta\lambda<\Lambda_2^{1/4}\), where the equation resembles a diffusion equation with non-homogenous diffusivity \(\mathcal
{K}(r)\) which depends on \(\lambda(r)\) for \(0\leq r\leq 2\sqrt{2}\Lambda_2^{1/4}\) and decays exponentially for \(r\geq2\sqrt{2}\Lambda_2^{1/2}\),
\begin{equation}\label{diffusive}
    \partial_t\delta\lambda=\begin{cases}
        \mathcal
{K}(r)~\partial_r^2\delta\lambda&\text{for}~0\leq r\leq 2\sqrt{2}\Lambda_2^{1/4},\\
-\tilde{\eta}~\delta\lambda&\text{for}~r\geq2\sqrt{2}\Lambda_2^{1/2}.
    \end{cases}
\end{equation}
For \(\lambda(r)\) satisfying the static configuration in Figure \ref{fig:nz_LambdaVariation} we find that \(\mathcal
{K}(r)>0\) in its defined region as  implied by Eq. \eqref{diffusivity} below
\begin{equation}\label{diffusivity}
    \mathcal
{K}(r)=\tilde{\eta}\left(r^2-8\sqrt{\Lambda_2}\right)\left(\frac{ -128\Lambda_2  -r^4+24 r^2\sqrt{\Lambda_2}}{256\Lambda_2 }\right).
\end{equation}
Thus, since the \(\delta\lambda\) is diffusive, the Skyrmion from the static minimum energy configuration we found previously are stable under small first-order perturbation. We demonstrate this stabilization to the unperturbed solution in Figure \ref{fig:DynamicsMagnetization}, where we plot the \(\hat
{z}\) component of the average magnetization, \(\vec{M}=\left<\vec{n}\right>\), against time for the case of initial perturbation \(\delta\lambda(0,r)=\delta\lambda_0~(2\sqrt{2}\Lambda_2^{1/4}-r)/2\sqrt{2}\Lambda_2^{1/4}\), \(\delta\lambda_0\) is a constant. The initial perturbation profile satisfies the continuity condition of \(\lambda(r)\) at all \(r\). We can see that the magnetization saturates to its unperturbed value at \(t\rightarrow\infty\).

The total energy of the system satisfies the following Bogomolnyi bound \cite{Bogomolny:1975de,1989JETPBogdanov},
\begin{eqnarray}
   H&\geq& 32\pi\kappa_0\sqrt{\Lambda_2}\int^0_R\frac{(\lambda/r)^2}{\left((\lambda/r)^2+1\right)^{5/2}}\frac{d}{dr}\left(\frac{\lambda}{r}\right)~dr \nonumber\\&=&  \frac{32\pi\kappa_0}{3}\sqrt{\Lambda_2}.\label{LowBound}
\end{eqnarray}
The existence of this lower bound for the energy strengthens the argument for the stability which implies that the Skyrmion cannot disperse into vacuum configuration \((H=0,~ \vec{n}(x,y)=\hat{z})\) spontaneously. One can observe that the minimum energy in Eq. \eqref{LowBound} is proportional to \(\Lambda_2^{1/2}\) which implies that the minimum energy is proportional to \(r_s^2\). Such property in a two-dimensional system is in agreement with the generalized arbitrary dimensional Skyrmion where the lowest possible energy is always proportional to its volume for \(Q\neq0\) \cite{Fadhilla:2021jiz}. The saturation of this energy bound happens if we take the invariant in \eqref{invar} to be zero, otherwise, the solution of \(\lambda(r)\) does not saturate this lower bound in general.
\section{\label{Conclusions}Conclusions and Outlook}
We introduce an effective model for topologically stable Skyrmions under a strong external perpendicular magnetic field through the Zeeman effect where the usual exchange interactions are negligible. The defining feature of the model is the Hamiltonian that uses the squared Skyrmion number density, which is shown to be relevant at the strong external magnetic field limit. It is shown in Section \ref{Result} that the minimum energy configuration of Hamiltonian \eqref{Hamiltonian} produces a skyrmion texture by using BP solutions with varying \(\lambda(r)\) as the ansatz. The solutions can be found in Figs. \ref{fig:BPSSkyrmionAbove} and \ref{fig:nz_LambdaVariation} for specific values of \(\Lambda_2\) and the general solutions possess the same qualitative features. These solutions are minimum energy configuration of Eq. \eqref{Hamiltonian} and are stable under small radial perturbations through LLG dynamics, given by Eq. \eqref{LLGEquation}. However, the stability under more general perturbation without axial symmetry is still an open problem for this model. The stabilization of the perturbed configuration to the minimum energy configuration is demonstrated in Fig. \ref{fig:DynamicsMagnetization} where we plot the average magnetization over time for several magnitudes of radial perturbation. We also analytically show that the total energy is bounded below in \eqref{LowBound} which is in agreement with the stable property we found through LLG dynamics. Such a lower bound is typical in topologically stable solitons \cite{Bogomolny:1975de,Manton:2004tk,Atmaja:2015umo}.

Although the model has been shown to produce a consistent result with the usual properties of magnetic Skyrmions, the microscopic origin of the \(q^2\) interaction is still unknown and requires more rigorous analyses. This is because this interaction term comes from a rather geometrical argument from the construction of generalized harmonic maps. Because the term is unique for two-dimensional systems and behaves as \(O(\vec{n}^6)\) an early guess for the origin of this interaction could be from the crystalline anisotropy \cite{yu2021magnetic,stancil2009spin,1990PhRvB..4111919D,rana2019towards}. Furthermore, the study on the more general model where the usual interaction terms are not neglected can also shed some light on how the Skyrmion textures are produced, especially when the thermal effect is also considered. However, the resulting dynamical equations are highly non-linear due to the incorporation of \(q^2\) term, and they require techniques beyond the ones we used in this work. We are going to address these problems in future works.

\section*{Acknowledgements}
E. S. F. would like to acknowledge the support from Badan Riset dan Inovasi Nasional through the Post-Doctoral Program 2024. The authors would like to thank A.R.T. Nugraha for discussions on topics related to two-dimensional systems.

\bibliographystyle{acm} 
\bibliography{main}

\appendix
\section{Generalized Skyrme Model}
\subsection{Manton's construction of Skyrme Model}
In this section, we are going to introduce the construction of the Skyrme model which is viewed as a generalized harmonic map. The ideas provided here are explained intuitively and a more rigorous description of the construction can be found in \cite{manton1987,Arthur:1996ia,Manton:2004tk}. Since the spins satisfy \(\vec{n}\cdot\vec{n}=1\), then \(\vec{n}_{m,n}\) is effectively a map from two-dimensional Euclidean space to a 2-sphere, \(\vec{n}:\mathbb{R}^2\rightarrow S^2\). We can define a Jacobi matrix since both the base and target space are smooth, whose components are \(J^\alpha_i=\partial y^\alpha/\partial x^i\), where \(y^\alpha\) are coordinates in \(S^2\) and \(x^i\) are coordinates in \(\mathbb{R}^2\). The coordinate system on \(S^2\) itself takes value in an open subset \(\mathcal{U}\in\mathbb{R}^2\), thus we can define a strain tensor, \(D\equiv JJ^T\), which can be imagined as the measure of how much the map, \(\vec{n}\), "deforms" the coordinates in target space from the base space on each direction. In terms of \(\vec{n}\), we can express the components of \(D\) as
\begin{equation}
    D_{ij}=\partial_i\vec{n}\cdot\partial_j\vec{n},
\end{equation}
which can also be viewed as the pull-back of the metric tensor on the target space to the base space, \(\mathbb{R}^2\). 

According to Manton's work on the geometry of the Skyrme model \cite{manton1987}, we can view the map defined above as a harmonic map, which satisfies a certain variational minimization of energy functional, known as the Dirichlet functional. The generalized version of the Dirichlet functional contains all of the invariants of \(D\), \(\text{Tr}[D]\) and \(\text{det}[D]\), namely
\begin{equation}
     \mathcal{E}=\int d^2x ~\left[\partial_i \vec{n}\cdot \partial^i \vec{n}+\kappa\left(\partial_x \vec{n}\times \partial_y \vec{n}\right)\cdot\left(\partial_x \vec{n}\times \partial_y \vec{n}\right)\right],
\end{equation}
where \(\kappa\) is the relative coupling between the two terms.

The second term above happens to be proportional to the squared Skyrmion number density, \(q^2\).
We can see that in this generalized model, the second term above contributes to the energy functional in a similar way to the Heisenberg term, \(\partial_i \vec{n}\cdot \partial^i \vec{n}\). Thus, the Skyrmion number (or topological charge) can also show up explicitly in the Hamiltonian.
\subsection{Discretization and magnetic lattice system}\label{Discret}
 Using the discretization \eqref{DiscreteSkyrmeNumber}, we have
 \begin{eqnarray}
     Q&=&\sum_{m,n}^{M,N} q_{m,n}~a^2.
 \end{eqnarray}
 For small \(h_x\) and \(h_y\) we can approximate
    \(h_x\partial_x\vec{n}(x,y)\simeq \vec{n}_{m+1,n}-\vec{n}_{m,n}\) and \(h_y\partial_y\vec{n}(x,y)\simeq \vec{n}_{m,n+1}-\vec{n}_{m,n}\).
As such, for the case of small \(a\), the topological charge goes back to the usual expression, \(q=\vec{n}\cdot\left(\partial_x\vec{n}\times\partial_y\vec{n}\right)/(4\pi)\) and \(Q=\int q ~d^2x.\) From the Hamiltonian Eq. \eqref{Hamiltonian}, we can see that the Zeemann term scales as \(a^2\) and Skyrme term scales as \(a^{-2}\) are important for Derrick's stability because this way we can ensure that there exists positive lower bound of the energy such that for scaling \(a\rightarrow \mathcal{K} a\) we have \(dH/d\mathcal{K}=0\) and \(d^2H/d\mathcal{K}^2>0\) at a certain positive \(\mathcal{K}\). The scaling analyses can also be used to estimate the energy scale at which our proposed model is relevant. Under \(a\rightarrow \mathcal{K} a\), the general Hamiltonian scales as
\begin{equation}
    H=\mathcal{K}^2H_Z+H_H+\mathcal{K}H_{DM}+\frac{H_q}{\mathcal{K}^2},
\end{equation}
 Let \(\mathcal{K}_{S}\) solves \(dH/d\mathcal{K}=0\). We found that \(d^2H/d\mathcal{K}^2>0\) for \(H_q\geq0\) and \(H_Z>0\), which implies that the system is stable regardless of \(\mathcal{K}_{S}\), according to the Derrick's theorem. For the case of \(H_q=0\), \(H_{DM}\) must be negative in order to have positive \(\mathcal{K}_S\), which is satisfied by the conventional Skyrmion solutions. For \(H_q>0\), the case with \(H_{DM}=0\) is allowed and positive \(\mathcal{K}_S\) can be produced from competition between \(H_Z\) and \(H_q\).
\section{The Skyrmion texture}
 \subsection{Variational method and Skyrmion solutions}
We can see that the total energy of this model at continuum limit, Eq. \eqref{EnergyBPSModel}, does not depend on the helicity. Then, we proceed to find the radial profile of the Skyrmion by first introducing a new dimensionless profile \(\xi=\lambda/r\) and followed by finding the solution of the EL equation of \eqref{EnergyBPSModel} with respect to \(\xi\). The EL equation for \(\xi\) gives us an invariant,
\begin{eqnarray}\label{invar}
    \frac{d}{dr}\left(16\Lambda_2\frac{\xi^2}{(1+\xi^2)^4}\frac{\xi'{}^2}{r^2}-\frac{\xi^2}{1+\xi^2}\right)=0.
\end{eqnarray}
For the special case where this invariant is zero
we have a first-order equation with two possible solutions, namely the trivial solution \(\xi_{\text{triv}}=0\) (this also implies \(\lambda_{\text{triv}}=0\)) and the non-trivial solution. As seen from the equation that is quadratic in \(\xi'\), we have two possibilities for the non-trivial solution from the signature of the equation. The positive signature corresponds to the Skyrmion solution, while the equation with negative signature has the opposite behaviour which corresponds to an anti-Skyrmion. Solving the first-order equation above for the non-trivial solution, gives us Eq. \eqref{ProfileLambda}. In this BP ansatz, the topological charge is equal to one,
\begin{equation}
    Q=\frac{1}{4\pi}\int_0^R \left(4\lambda\frac{\lambda-r\lambda'}{(\lambda^2+r^2)^2}\right) r~d\varphi~dr=2\int_0^\infty\frac{\xi}{(1+\xi^2)^2}d\xi=1.
\end{equation}

\subsection{Radial Stability}
Assume that the unperturbed Skyrmion profile satisfies \eqref{ProfileLambda}. For \(r>2\sqrt{2}\Lambda_2^{1/4}\) the solution is trivial, namely \(\delta\lambda=0\) and \(\delta\gamma\) is an arbitrary constant. Hence, we just need to focus on the behaviour of \(\delta\lambda\) in the region \(0\leq r\leq 2\sqrt{2}\Lambda_2^{1/4}\). The  first-order perturbed LLG equation implies that
\begin{eqnarray}
\partial_t\delta\gamma&=&0,\\
    \partial_t\delta\lambda&=&\tilde{\eta}\left(\bar{r}^2-8\right)\left[\left(\frac{35 \bar{r}^4-760 \bar{r}^2+3584   }{256   (16
    -\bar{r}^2)}\right)\delta\lambda\right.\nonumber\\&&\left.+\left(\frac{ \left(-384  -13 \bar{r}^4+184  \bar{r}^2\right)}{256  \bar{r}}\right)\partial_r\delta\lambda\right.\nonumber\\&&\left.+\left(\frac{ \left(-128  -\bar{r}^4+24\bar{r}^2\right)}{256 }\right)\partial_r^2\delta\lambda\right],\label{PerturbedLambda}
\end{eqnarray}
with \(\bar{r}=r/\Lambda_2^{1/4}\). If the scale of the radial perturbation is much smaller than the length scale from the effective coupling, i.e. \(\delta\lambda<<\Lambda_2^{1/4}\), then the term with \(\delta\lambda''\) is the dominant term under this assumption. As such, we can approximate \eqref{PerturbedLambda} as a diffusion equation with a non-homogenous diffusivity, given by Eq. \eqref{diffusive}.

For a more general approach where the scale of the perturbation is not assumed, the problem of stability can be reduced into a Sturm-Liouville eigenvalue problem. Equation \eqref{PerturbedLambda} is linear in \(\delta\lambda\), hence it is separable, \(\delta\lambda=\psi_t(t)\psi_r(r)\). The solution of \(\psi_t\) is straightforward, namely \(\psi_t=e^{-\omega t}\), where \(\omega\) is the stability parameter which satisfies the eigenvalue equation
\begin{eqnarray}
    -\omega\psi_r&=&\tilde{\eta}\left(\bar{r}^2-8\right)\left[\left(\frac{35 \bar{r}^4-760 \bar{r}^2+3584   }{256   (16
    -\bar{r}^2)}\right)\psi_r\right.\nonumber\\&&\left.+\left(\frac{ \left(-384  -13 \bar{r}^4+184  \bar{r}^2\right)}{256  \bar{r}}\right)\psi_r'\right.\nonumber\\&&\left.+\left(\frac{ \left(-128  -\bar{r}^4+24\bar{r}^2\right)}{256 }\right)\psi_r''\right].
\end{eqnarray}
For continuous perturbation, this equation should satisfy two boundary conditions, namely \(\psi_r(0)=0\) such that \(\vec{n}\) is continuous at the origin, and \(\psi_r(\bar{r}=2\sqrt2)=0\) such that value of \(\delta\lambda\) inside \(r\leq2\sqrt{2}\Lambda_2^{1/4}\) connected to the value at \(r>2\sqrt{2}\Lambda_2^{1/4}\). If all the solutions of this eigenvalue problem have \(\omega>0\) then the Skyrmion texture given by \eqref{ProfileLambda} is stable under continuous first-order perturbation. This open problem requires a more rigorous approach and will be addressed in future works dedicated solely to stability analysis in a more general manner.







\end{document}